\documentclass[%
 aip,
rsi,%
 amsmath,amssymb,
 reprint,%
]{revtex4-1}

\usepackage{xcolor}
\usepackage{graphicx}
\usepackage{dcolumn}
\usepackage{bm}
\usepackage{nicefrac}

\begin{document}

\preprint{AIP/123-QED}

\title[Rodr\'{i}guez et al.]{Higher order theory of quasi-isodynamicity near the magnetic axis of stellarators}

\author{E. Rodr\'{i}guez}
 \altaffiliation[Email: ]{eduardo.rodriguez@ipp.mpg.de}

\author{G. G. Plunk}%
\affiliation{%
Max Planck Institute for Plasma Physics, Wendelsteinstrasse 1, 17491 Greifswald, Germany
} 

\date{\today}

\begin{abstract}
The condition of quasi-isodynamicity is derived to second order in the distance from the magnetic axis.  We do so using a formulation of omnigenity that explicitly requires the balance between the radial particle drifts at opposite bounce points of a magnetic well.  This is a physically intuitive alternative to the integrated condition involving distances between bounce points, used in previous works.  We investigate the appearance of topological defects in the magnetic field strength (``puddles'').  A hallmark of quasi-isodynamic fields, the curved contour of minimum field strength, is found to be inextricably linked to these defects.  Our results pave the way to constructing solutions that satisfy omnigenity to a higher degree of precision, and also to simultaneously consider other physical properties, like shaping and stability.
\end{abstract}

\maketitle

\section{\label{sec:intro} Introduction}
The stellarator\cite{spitzer1958,boozer1998,Helander2014} is an attractive magnetic confinement concept to achieve controlled thermonuclear fusion. Its flexibility and generality make it unique in its ability to confine plasmas avoiding common shortcomings of other confinement designs\cite{schuller1995,boozer1998}. However, to achieve good confinement, the field must be optimised\cite{mynick2006}. Omnigeneous stellarators\cite{bernardin1986,cary1997,hall1975,helander2009,landreman2012,Helander2014} are one such class of optimised stellarators, one that minimise the rapid loss of particles. Quasisymmetric stellarators\citep{boozer1983,nuhren1988,rodriguez2020} are a particular subclass. 
\par
In omnigeneous stellarators, in the absence of collisions, particles do not to drift away from the device, on average, which avoids the fast loss of particles. This property is achieved only when the magnitude of the magnetic field satisfies certain conditions\cite{landreman2012}. Even with these constraints provided, the class of approximately omnigeneous stellarators has a significant freedom compared to the more restrictive case of a tokamak\cite{mukhovatov1971,wessonTok}. This makes the design of stellarators versatile, but necessarily complicates their systematic study. 
\par
A possible way to approach the study of omnigeneous stellarators is to consider the behaviour of these configurations near their magnetic axes. This simplifying near-axis expansion approach\cite{mercier1962,Solovev1970,garrenboozer1991a,landreman2019,jorge2020a,rodriguez2020i} has been a recurrent analytic tool in the study of stellarators. In this paper we use it to present the consequences of omnigenity on the magnetic field magnitude near the magnetic axis. We specialise to the concept of quasi-isodynamicity, continuing the work initiated in [\onlinecite{plunk2019}].
\par
We do so by first presenting a physically intuitive definition of omnigenity in Section~\ref{sec:omnDef}. In Sec.~\ref{sec:QInae} we proceed to assess quasi-isodynamicity near the axis,. There, the property of quasi-isodynamicity is reduced to symmetry properties on the asymptotic form of $|\mathbf{B}|$. Section~\ref{sec:topDefects} treats the issues of pseudosymmetry and the possibility of topological defects in $|\mathbf{B}|$. Section~\ref{sec:conclusions} closes with a brief summary and conclusions.

\section{omnigenity condition} \label{sec:omnDef}
We define \textit{omnigenity} as the condition on the magnetic field such that the bounce average radial drift of trapped particles vanishes. This notion goes back to the seminal work by Hall \& McNamara\cite{hall1975}. Here we present one particularly physical form of the concept.
\par
Consider a magnetic field with nested toroidal flux surfaces labelled by the toroidal flux $2\pi\psi$. On every surface, the magnetic field satisfies $\mathbf{B}\cdot\nabla\psi=0$ and thus can be conveniently written in the Clebsch form\cite{dhaeseleer2012} as $\mathbf{B}=\nabla\psi\times\nabla\alpha$, where $\alpha$ is the field line label $\alpha=\theta-\iota\varphi$. Here $\varphi$ and $\theta$ are the toroidal and poloidal Boozer angles\cite{boozer1981} respectively, and $\iota$ is the rotational transform\cite{Helander2014}. The use of Boozer coordinates is possible as we consider $\mathbf{B}$ to be in equilibrium, satisfying $\mathbf{j}\times\mathbf{B}=\nabla p$, where $\mathbf{j}=\nabla\times\mathbf{B}$ is the current density and $p$ is the scalar pressure. It will be convenient to think of the magnitude of the magnetic field, $B$, as a separate function of the two angular coordinates at each magnetic flux surface.
\par
\begin{figure*}
    \centering
    \includegraphics[width=0.9\textwidth]{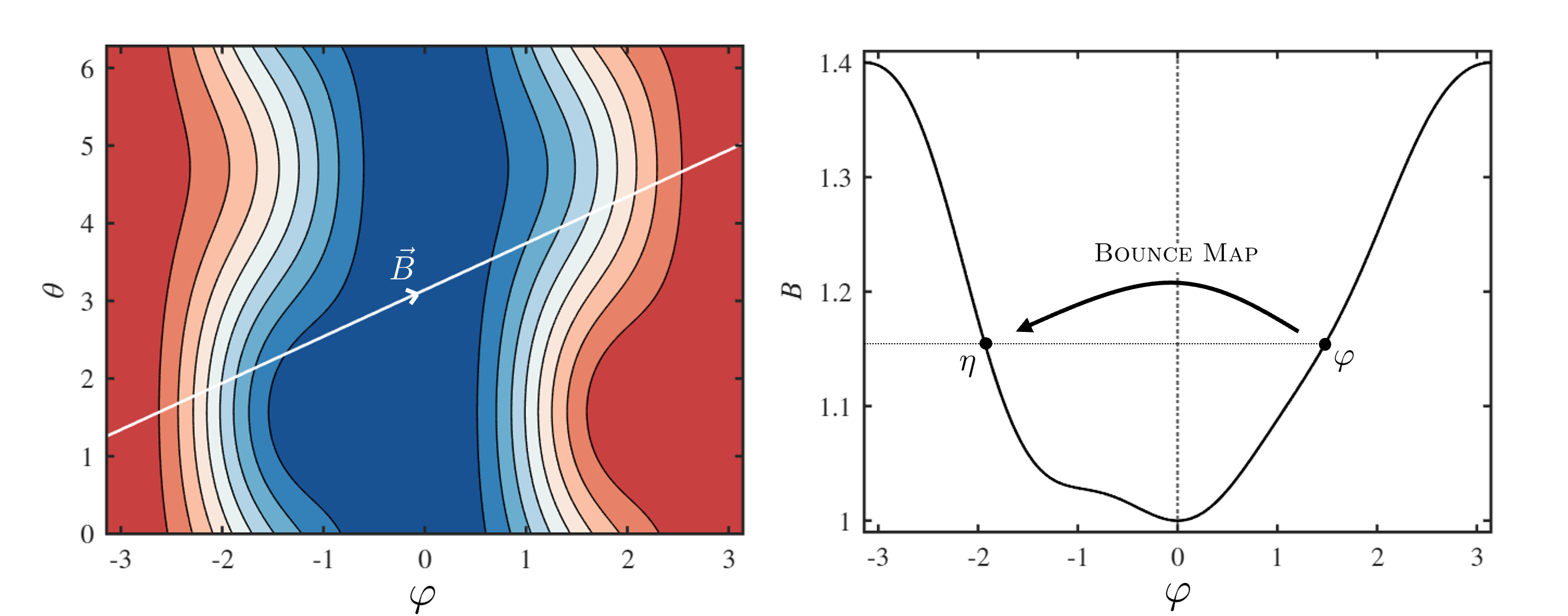}
    \caption{\textbf{Bounce map definition.} The plot on the left is an example of contours of magnetic field magnitude on a flux surface in Boozer coordinates $(\theta,\varphi)$. The white line is an example of a magnetic field line, and $|\mathbf{B}|$ along the field line is shown on the plot to the right. There the meaning of the bounce map $\eta$ is illustrated.}
    \label{fig:my_label}
\end{figure*}

With this set-up, we are interested in studying the motion of trapped particles. Charged particles, to leading order in guiding centre theory\cite{littlejohn1983}, move along magnetic field lines as they gyrate about them. They do so by conserving energy, $W$, as well as the first adiabatic invariant, the magnetic moment $\mu$. This conservative motion is what leads to the notion of \textit{trapped particles} (also known as \textit{bouncing particles}). From the definition of $\mu=W_\perp/B$, where $W_\perp$ is the kinetic energy of the particle in the direction normal to the field, particles with a particular value of $W$ and $\mu$ are excluded from regions for which $B>W/\mu$. As particles approach such regions they are forced back. The result is particles that bounce back and forth along field lines between points at which $B=W/\mu\stackrel{\cdot}{=}1/\lambda$. Those points are called \textit{bouncing points}, and come in pairs dictated by $|\mathbf{B}|$. Formally these are related through a bouncing function $\eta(\varphi, \theta,\psi)$ (see Fig.~\ref{fig:my_label}). We define this function to be the toroidal angle value of a bouncing point paired to one at $\varphi$ along a field line $\alpha$, that is, the point with the same value of $B$ on the other side of the well. This should not be confused with $\eta$ in previous work by Cary \& Shasharina\cite{cary1997} and Plunk et al.\cite{plunk2019}. The parameter $\lambda$ can be interpreted as defining different \textit{classes} of bouncing particles. 
\par
The main issue with trapped particles is that they do not only bounce, but they are also subject to guiding centre drifts. Neglecting electric fields and focusing on magnetic field inhomogeneity, from guiding centre theory (and taking a positive charge of 1) the drift velocity is given by\cite{littlejohn1983,Alfven1940},
\begin{equation}
    \mathbf{v}_D=-\frac{2W_\parallel}{B^2}\kappa\times\mathbf{B}-\frac{\mu}{B^2}\nabla B\times\mathbf{B},
\end{equation}
which upon use of force balance and the relation $\kappa\times\mathbf{B}=[\nabla p\times\mathbf{B}+B\nabla B\times\mathbf{B}]/B^2$ yields,
\begin{equation}
    \mathbf{v}_D\cdot\nabla\psi=\frac{(2-\lambda B)W}{B^3}\mathbf{B}\times\nabla B\cdot\nabla\psi.
\end{equation}
To assess the implications of this drift on the trapped particle confinement, we must consider the average of $\mathbf{v}_D\cdot\nabla\psi$ as the particles move along the field line. As $\mathrm{d}t=\mathrm{d}l/v_\parallel$, with $l$ the length along the field line, for each class of trapped particles the net drift per bounce, $\Delta\psi$, is given by (normalising mass to 1)
\begin{equation}
    \Delta\psi=\sqrt{2W}\oint_{\alpha,\psi}\frac{1-\lambda B/2}{\sqrt{1-\lambda B}}\frac{\mathbf{B}\times\nabla B\cdot\nabla\psi}{B^3}\mathrm{d}l, \label{eqn:deltaPsi1}
\end{equation}
where the integral is taken between two bounce points.
\par
The integral in Eq.~(\ref{eqn:deltaPsi1}) has a certain symmetry through $B$, as the integral limits show. Thus, it is convenient to rewrite it using $B$ explicitly as a parameter along field lines. The change in the integral measure is straightforward, as $\mathrm{d}l=\mathrm{d}B/\mathbf{b}\cdot\nabla B$. We must however consider the right and left parts of the integral separately. That is, with the well-minimum at $l=0$,
\begin{align*}
   \oint_{\alpha,\psi}\mathrm{d}l&=2\int_{l_L}^{l_R}\mathrm{d}l=2\left[\int_{l_L}^{0}+\int_{0}^{l_R}\right]\mathrm{d}l=\\
   &=2\left[\int_{1/\lambda}^{B_\mathrm{min}}+\int_{B_\mathrm{min}}^{1/\lambda}\right]\frac{\mathrm{d}B}{\mathbf{b}\cdot\nabla B}, \\
\end{align*}
where $l_L$ and $l_R$ are the left and right bouncing points. For the terms in the integrand of Eq.~(\ref{eqn:deltaPsi1}) that only depend on $B$ and $\lambda$ (the first factor $f(\lambda,B)=(1-\lambda B/2)/(B^2\sqrt{1-\lambda B})$), the two halves of the integral are identical (up to the sign coming from flipping the limits of the integral). For the other factors in the integrand the integrals are generally different, as they involve different portions of the field line. Introducing $\gamma=\mathrm{sign}(\mathbf{B}\cdot\nabla B)=\pm1$ to indicate which half of the well is being considered, Eq.~(\ref{eqn:deltaPsi1}) may be written succintly as\cite{plunk2019},
\begin{equation}
    \Delta\psi=2\sqrt{2W}\int_{B_\mathrm{min}}^{1/\lambda}f(\lambda,B)\sum_\gamma\gamma Y \mathrm{d}B, \label{eqn:deltaPsi2}
\end{equation}
where,
\begin{equation}
    Y=\frac{\nabla \psi\times\mathbf{B}\cdot\nabla B}{\mathbf{B}\cdot\nabla B}, \label{eqn:Ydef}
\end{equation}
and it is a function of $\psi$, $\alpha$, $B$ and $\gamma$. In the event of multiple wells this splitting of the integral may be easily generalised to include all appropriate segments along the field line. This is laborious but straightforward, and thus we focus on the situation of a single well for simplicity.
\par
Because the $B$-symmetric factor $f(\lambda,B)$ is positive, for the integral to vanish for all $\lambda$, it must be that,\footnote{It is simple to picture why the integrand of Eq.~(\ref{eqn:deltaPsi2}) must vanish. For large $\lambda$, the integral only samples the bottom of the well, where this must be true. As $\lambda$ decreasees the range of the integral increases, but because all the previously included values must vanish, then this must hold everywhere. It corresponds to an invertible integral transform.}
\begin{equation}
    \sum\gamma Y=0, \label{eqn:QIsumDef}
\end{equation}
for all $B$ values in the well. That is, the geometric quantity $Y$, Eq.~(\ref{eqn:Ydef}), must be the same on each side of the well. Physically, this amounts to requiring the radial drift on each side of the well (generally non-zero) to exactly cancel out so that there is no net drift.
\par
It is most convenient to express $Y$ as a function of Boozer coordinates explicitly, so Eq.~(\ref{eqn:QIsumDef}) is,
\begin{equation}
    Y(\psi,\theta,\varphi)-Y[\psi,\theta+\iota\Delta\varphi,\eta(\psi,\theta,\varphi)]=0, \label{eqn:omnDefY}
\end{equation}
where $\Delta\varphi=\eta(\varphi, \theta, \psi)-\varphi$ is, by the definition of $\eta$, the angular distance between bouncing points. This, Eq.~(\ref{eqn:omnDefY}), we shall take as the formal definition of omnigenity. 
\par

\section{Near-axis approach to QI} \label{sec:QInae}
We now study the property of omnigenity close to the magnetic axis of a toroidal magnetic configuration. This requires an asymptotic description of the magnetic field, and an appropriate handling of the omnigenity property in Eq.~(\ref{eqn:omnDefY}).

\subsection{Magnetic field magnitude}
Let us start by considering the form of the magnetic field magnitude perturbatively in the distance from the magnetic axis. Defining the pseudo-radial coordinate $\epsilon=\sqrt{2\psi/\bar{B}}$, where $\bar{B}$ is a reference magnetic field, we may write the magnetic field asymptotically to second order, following Garren \& Boozer\cite{garrenboozer1991a}, as
\begin{equation}
    B(\psi,\theta,\varphi)=B_0(\varphi)+\epsilon B_1(\theta,\varphi)+\epsilon^2B_{2}(\theta,\varphi), \label{eqn:Bform}
\end{equation}
\begin{subequations}
\begin{equation}
    B_1(\theta,\varphi)= B_0(\varphi)d(\varphi)\cos[\alpha+\nu(\varphi)], \label{eqn:B1}
\end{equation}
\begin{multline}
    B_2(\theta,\varphi)=B_{20}(\varphi)+B_{2c}(\varphi)\cos2[\alpha+\nu(\varphi)]
    +\\
    +B_{2s}(\varphi)\sin2[\alpha+\nu(\varphi)]. \label{eqn:B2}
\end{multline}
\end{subequations}
We have taken the liberty here of redefining the argument of the cosines and sines. Here $\alpha=\theta-\iota\varphi$ denotes the field-line label, and will prove to be a helpful form of writing $|\mathbf{B}|$.
\par
The expression in Eq.~(\ref{eqn:Bform}) presents two key features of the problem.  The first is the coupling of the harmonics of the field and the powers of $\epsilon$, necessary to enforce analyticity at the magnetic axis. In second place, the magnetic field on axis is generally a function of the toroidal angle. We will assume $B_0$ to have such a non-trivial $\varphi$ dependence, which amounts to specialising to quasi-isodynamic (QI) configurations within the class of omnigeneous fields. These correspond to optimised omnigenous configurations with poloidally closed $|\mathbf{B}|$ contours. This choice necessarily excludes the two other subclasses of omnigenous fields, those with toroidally and helically closed contours, as they can only be achieved near the magnetic axis if quasisymmetry is also satisfied.\cite{plunk2019} As in the treatment leading to Eq.~(\ref{eqn:omnDefY}), for simplicity, we shall assume there to be a single magnetic well, with its minimum located at $\varphi=0$.

\subsection{Radial drift measure $Y$}
Once we have the magnetic field magnitude in the form of Eq.~(\ref{eqn:Bform}), we are in a position to evaluate the radial drift measure $Y$. We must (i) first express Eq.~(\ref{eqn:Ydef}) in Boozer coordinates, and then (ii) consider its expansion in $\epsilon$.
\par
The first step is straightforward. It suffices to recognise that the magnetic differential operator $\mathbf{B}\cdot\nabla$ can be written as $\mathcal{J}^{-1}(\partial_\varphi+\iota\partial_\theta)$, where the partial derivatives are taken with respect to the Boozer set $\{\psi,\theta,\varphi\}$ and $\mathcal{J}$ is the coordinate Jacobian.\cite{boozer1981} In Boozer coordinates $\mathbf{B}=I\nabla\theta+G\nabla\varphi+B_\psi\nabla\psi$, where $I$ and $G$ are flux functions representing toroidal and poloidal currents, which then lend to,
\begin{equation}
    Y=\frac{(I\partial_\varphi-G\partial_\theta) B}{(\partial_\varphi+\iota\partial_\theta)B}.
\end{equation}
Perturbatively, using Eq.~(\ref{eqn:Bform}), and expanding the flux functions as $G=G_0+\epsilon^2 G_2+\dots$ (and equivalently for $I$ and $\iota$),
\begin{equation}
    (I\partial_\varphi-G\partial_\theta)B = -\epsilon G_0\partial_\theta B_1+\epsilon^2(I_2B_0'-G_0\partial_\theta B_2)+O(\epsilon^3), \label{eqn:vddotdpsi}
\end{equation}
\begin{equation}
    (\partial_\varphi+\iota\partial_\theta)B = 
    B_0'+\epsilon(\partial_\varphi+\iota_0\partial_\theta)(B_1+\epsilon B_2)+O(\epsilon^3), \label{eqn:BdotB}
\end{equation}
where we considered $I_0=0$ (that is, no current singularity on the axis\cite{garrenboozer1991a,landreman2019}). 
\par
Putting everything together we get $Y=\epsilon Y^{(1)}+\epsilon^2 Y^{(2)}+\dots$,
\begin{subequations}
    \begin{gather}
        Y^{(1)}= -\frac{G_0}{B_0'}\partial_\theta B_1, \label{eqn:Y1}
    \\
        Y^{(2)}=I_2+\frac{G_0}{(B_0')^2}\partial_\theta B_1(\partial_\varphi+\iota_0\partial_\theta)B_1-\frac{G_0}{B_0'}\partial_\theta B_2. \label{eqn:Y2}
    \end{gather}
\end{subequations}
It is paramount to the asymptotic construction to assume that $B_0'\neq 0$, as the asymptotic division by $\mathbf{B}\cdot\nabla B$ in Eqs.~(\ref{eqn:Y1})-(\ref{eqn:Y2}) requires $\epsilon\ll B_0'$. Although the asymptotic procedure will generally apply given our quasi-isodynamic assumption\cite{plunk2018}, it will fail in regions where $B_0'\sim 0$. These points are special even in a non-asymptotic sense. Barely and deeply trapped particles spend an infinite amount of time at these locations, and thus if these classes are to be confined (as they should in an omnigeneous field), the radial drift, Eq.~(\ref{eqn:vddotdpsi}), must vanish there. This precise property is known as \textit{pseudosymmetry}\cite{mikhailov2002,skovoroda2005}, and it requires $\nabla B\times\mathbf{B}\cdot\nabla\psi=0$ wherever $\mathbf{B}\cdot\nabla B=0$. As a result, all contours of $B$ over flux surfaces must share the same topology\cite{landreman2012}. Because the asymptotics of Eq.~(\ref{eqn:omnDefY}) fail in an asymptotically small region near the turning points, explicitly imposing pseudosymmetry is a way of ensuring that omnigeneity is not being spoiled there. Because of its global topological implications, imposing this condition explicitly will naturally lead to notions on the topology of the contours of $|\mathbf{B}|$. The details of asymptotically studying the pseudosymmetry condition are presented in Appendix \ref{sec:appPseudo}, and the main results of this are discussed in Sec.~\ref{sec:2ndOrderQI}.

\subsection{Bounce map}

As we perturb the magnetic field magnitude going from one order $\epsilon$ to the next, the shape of $B$ along field lines changes. With it the bounce map $\eta$ also changes. To describe it accurately we must define $\eta$ more precisely as a function of all three Boozer coordinates such that,
\begin{equation}
    B[\psi,\theta-\iota(\varphi-\eta),\eta]=B(\psi,\theta,\varphi), \label{eqn:defEtaMap}
\end{equation}
excluding the trivial solution $\eta=\varphi$ (except at minima where this is always satisfied). That is, at a given flux surface and magnetic field line, the map takes a point $\varphi$ on one side of the well to the value at other side with the same $B$ (see Fig.~\ref{fig:my_label}). 
\par
From this definition it is clear that, if we perturb $B$, then the function $\eta$ will have to adjust accordingly. Thus, it is appropriate to think of the map $\eta$ also perturbatively, $\eta=\eta_0(\varphi)+\epsilon \eta_1(\varphi,\theta)+\dots$. Substituting this and Eq.~(\ref{eqn:Bform}) into Eq.~(\ref{eqn:defEtaMap}), we obtain the following upon collecting terms. To order $\epsilon^0$,
\begin{subequations}
    \begin{equation}
        B_0[\eta_0(\varphi)]=B_0(\varphi), \label{eqn:eta0}
    \end{equation}
    and to order $\epsilon$,
    \begin{equation}
        \eta_1(\theta,\varphi)=\frac{B_1(\theta,\varphi)-B_1[\theta+\iota_0\Delta\varphi_0,\eta_0(\varphi)]}{B_0'[\eta_0(\varphi)]}, \label{eqn:eta1}
    \end{equation}
\end{subequations}
where $\Delta\varphi_0=\eta_0(\varphi)-\varphi$. The former is a definition of the map $\eta_0$ based on the structure of the magnetic field along the magnetic axis. This will be a fundamental feature of the construction, as it defines asymptotically the trapped particle classes. It is important to note that it is $\theta$ independent, not by assumption but forced by the behaviour of $B$ on axis. We will need a few relations coming from this expression, including,
\begin{gather}
    \eta'_0(\varphi)=\frac{B_0'(\varphi)}{B_0'[\eta_0(\varphi)]}, \label{eqn:eta0pdB0} \\
    \eta_0'(\varphi)=\frac{1}{\eta_0'[\eta_0(\varphi)]},\label{eqn:eta0prime}
\end{gather}
Note that we have used $\eta_0[\eta_0(\varphi)] = \varphi$, i.e. $\eta_0$ is self-inverse, to derive Eqs.~(\ref{eqn:eta0pdB0})-(\ref{eqn:eta0prime}) from (\ref{eqn:eta0}); it also follows from these two that $\eta_0' = -1$ at minima and global maxima of $B_0$.
\par
The second equation, Eq.~(\ref{eqn:eta1}), describes the change in the bounce map as a result of the first order perturbation. If the change in $B$ to the right and the left of the well is not the same, then the map function changes. This is generally the case, even when stellarator symmetry is invoked.

\subsection{QI condition}
The conditions for omnigenity are obtained by bringing the expansion of the drift $Y$ and the bounce map together. 

\subsubsection{Leading order $O(\epsilon)$}
To leading order, the omnigenity condition is,
\begin{equation}
    Y^{(1)}(\theta,\varphi)-Y^{(1)}(\theta+\iota_0\Delta\varphi_0,\eta_0(\varphi))=0, \label{eqn:QIY1}
\end{equation}
which reduces to,
\begin{equation}
    \partial_\theta B_1(\theta,\varphi)=\eta_0'(\varphi)\partial_\theta B_1[\theta+\iota_0\Delta\varphi_0,\eta_0(\varphi)]. \label{eqn:B1omni}
\end{equation}
Note that Eq.~(\ref{eqn:B1omni}) can be written without the differentiation with respect to the poloidal angle, as $B_1$ has a vanishing $\theta$ average (see Eq.~(\ref{eqn:B1})). 
\par
Let us now use the explicit form in Eq.~(\ref{eqn:B1}) for $|\mathbf{B}|$. Plugging it into Eq.~(\ref{eqn:B1omni}) yields,
\begin{equation}
    d(\varphi)\sin [\alpha+\nu(\varphi)]=\eta_0'd[\eta_0(\varphi)]\sin (\alpha+\nu[\eta_0(\varphi)]).
\end{equation}
Using trigonometric identities, and requiring the condition to hold for all field lines (namely, for all $\alpha$), continuous functions $\nu$ and $d$ must satisfy, 
\begin{subequations}
    \begin{gather}
        d(\varphi)=\eta_0'(\varphi) d[\eta_0(\varphi)], \\
        \nu(\varphi)=\nu[\eta_0(\varphi)]. \label{eqn:omnNu}
    \end{gather} \label{eqn:omnCondeps}
\end{subequations}
These are precisely the same symmetry conditions obtained in [\onlinecite{plunk2019}] using the Cary-Shasharina construction\cite{cary1997} (see also Appendix~\ref{sec:appCS}). The amplitude of the magnetic field correction $B_1$ must be, in a sense, odd about the minimum of $B_0$, while the function $\nu$ should be even.
\par
Following the parity condition, $d = 0$ at the minimum of $B_0$. The same argument follows for the global maximum of $B_0$. In the scenario of multiple wells, the QI condition in Eqs.~(\ref{eqn:omnCondeps}) do not appear to require the vanishing of $d$ at maxima which are not global. We must however be careful, as the asymptotic description fails in the neighbourhood of the turning points. To enforce omnigeneity in that region, we must enforce pseudosymmetry to first order: that is, limit the radial drift of barely and deeply trapped particles to be $O(\epsilon^\alpha)$, with $\alpha>1$. As shown in Appendix~\ref{sec:appPseudo}, this requires $B_1$ to vanish at \textit{all turning points} of $B_0$. At the minima enforcing Eqs.~(\ref{eqn:omnCondeps}) is enough. This is consistent with [\onlinecite{plunk2019}]. 
\par
It is noteworthy that, as recognised by [\onlinecite{plunk2019}], it is impossible to satisfy the QI conditions, Eqs.~(\ref{eqn:omnCondeps}), exactly. Doing so necessarily makes $|\mathbf{B}|$ non-periodic, following the secular behaviour of the field line label $\alpha$ (which at fixed $\theta$ gives $\alpha\rightarrow\alpha-2\pi\iota$ after one toroidal turn). Formally speaking, periodicity requires $\nu(2\pi)=\nu(0)+2\pi(\iota-N)$ for some integer $N$ (related to the self-linking number of the axis\cite{plunk2018,rodriguez2022phase}), a difference that must be zero for omnigeneous fields, Eq.~(\ref{eqn:omnNu}). Thus, omnigeneity must be broken at the (global) top of the wells. Although this may appear to negate any higher order study of omnigeneity, there are practical strategies to address this. For instance, one is free to define intervals in $\varphi$ around minima of $B_0$ where Eq.~(\ref{eqn:omnNu}) is, if desired, satisfied exactly, and thus all trapped particles residing in this interval behave in an omnigenous way, to that accuracy. In addition, the pseudosymmetry requirement reduces the losses from breaking omnigeneity near the top, as it minimises the magnitude of the radial drifts there. On a practical level, it is worth noting that approximately QI solutions found by numerical optimisation can deviate significantly from first order solutions, and it is certainly useful that such deviations manage to preserve omnigenity. We thus continue with the procedure to higher order, assuming Eqs.~(\ref{eqn:omnCondeps}) to be satisfied.

\subsubsection{Second order $O(\epsilon^2)$} \label{sec:2ndOrderQI}
At next order we find a term coming from the correction to the map,
\begin{multline}
     Y^{(2)}(\theta,\varphi)=Y^{(2)}(\theta+\iota_0\Delta\varphi_0,\eta_0(\varphi))+\\
     +\eta_1(\varphi)(\partial_2+\iota_0\partial_1) Y^{(1)}(\theta+\iota_0\Delta\varphi_0,\eta_0(\varphi)), \label{eqn:Y2Qi}
\end{multline}
A word of caution should be issued here with regards to the meaning of the notation in the second line. The partial derivatives $\partial_1$ and $\partial_2$ refer to derivatives with respect to, respectively, the first and second arguments of the function they act upon. The function is then evaluated for the arguments to the right. This term comes simply from Taylor expanding $Y^{(1)}$ under the perturbed $\eta$. Such a term is different from $(\partial_\varphi+\iota_0\partial_\theta)[Y_1(\theta+\iota_0\Delta\varphi_0,\eta_0(\varphi)]$, in which the partial derivatives act on the function in square brackets, and chain rule will be necessary. 
\par
The expression for $\eta_1$ was found in Eq.~(\ref{eqn:eta1}), and may be rewritten using the symmetry of $B_1$, Eq.~(\ref{eqn:B1omni}), as,
\begin{equation}
    \eta_1(\varphi)=\frac{B_1(\theta,\varphi)}{B_0'(\varphi)}[\eta_0'(\varphi)-1], \label{eqn:eta1div}
\end{equation}
where Eq.~(\ref{eqn:eta0pdB0}) was also used. 
\par
Then we need to evaluate,
\begin{align*}
    (\partial_2&+\iota_0\partial_1) Y^{(1)}(\theta+\iota_0\Delta\varphi_0,\eta_0(\varphi))=\\
    &=(\partial_\varphi+\iota_0\partial_\theta) [Y^{(1)}(\theta+\iota_0\Delta\varphi_0,\eta_0(\varphi))]\frac{1}{\eta_0'}=\\
    &=(\partial_\varphi+\iota_0\partial_\theta)Y^{(1)}(\theta,\varphi)\frac{1}{\eta_0'}=-\frac{G_0}{\eta_0'}(\partial_\varphi+\iota_0\partial_\theta)\left(\frac{\partial_\theta B_1}{B_0'}\right),
\end{align*}
where we used the chain rule for the first line, the QI condition Eq.~(\ref{eqn:QIY1}) for the second, and the form of $Y^{(1)}$ in Eq.~(\ref{eqn:Y1}) for the last. 
\par
Finally, we write $Y^{(2)}$ evaluated at the bounce $\eta_0(\varphi)$, using the same rationale as before,
\begin{multline}
    Y^{(2)}(\theta+\iota_0\Delta\varphi_0,\eta_0(\varphi))=\\
    I_2+\frac{G_0}{\eta_0'(B_0')^2}\partial_\theta B_1\left[(\partial_\varphi+\iota_0\partial_\theta)B_1-\frac{\eta_0''}{\eta_0}B_1\right]-\\
    -\frac{G_0}{B_0'}\eta_0'\partial_\theta B_2(\psi,\theta+\iota_0\Delta\varphi_0,\eta_0(\varphi)),
\end{multline}
where all the functions whose arguments not explicitly indicated are evaluated at $\varphi$ and $\theta$. 
\par
Putting all together into Eq.~(\ref{eqn:Y2Qi}), we can write
\begin{widetext}
\begin{equation}
    \partial_\theta\left[B_2(\theta,\varphi)-\eta'_0B_2(\psi,\theta+\iota_0\Delta\varphi_0,\eta_0(\varphi))-\left(1-\frac{1}{\eta_0'}\right)(\partial_\varphi+\iota_0\partial_\theta)\left(\frac{B_1^2}{2B_0'}\right)-\frac{\eta_0''}{(\eta_0')^2}\frac{B_1^2}{2B_0'}\right]=0\label{eqn:B2omni}
\end{equation}
\end{widetext}
The overall $\theta$ derivative is important, as it eliminates the $\theta$-independent part of $B_2$, Eq.~(\ref{eqn:B2}), meaning that the condition of QI does not impose any direct constraint on it. The other two components $B_{2c}$ and $B_{2s}$ do enter the equation, and upon requiring the condition to hold for all $\alpha$ as we did at first order,
\begin{widetext}
\begin{subequations}
    \begin{gather}
        B_{2s}(\varphi)=\eta_0'B_{2s}[\eta_0(\varphi)]-\frac{1}{2}\left(1-\frac{1}{\eta_0'}\right)\frac{B_0^2d^2}{B_0'}\nu', \\
        B_{2c}(\varphi)=\eta_0'B_{2c}[\eta_0(\varphi)]+\frac{1}{4}\left(1-\frac{1}{\eta_0'}\right)\partial_\varphi\left(\frac{B_0^2d^2}{B_0'}\right)+\frac{1}{4}\frac{\eta_0''}{(\eta_0')^2}\frac{B_0^2d^2}{B_0'},
    \end{gather} \label{eqn:omnB2f}
\end{subequations}
\end{widetext}
where, once again, the functions whose arguments are not indicated are evaluated at $\varphi$. The `symmetry' conditions at second order become more involved than the first order ones, involving the lower order choices. 
\par
To cast these conditions in a form that is closer to the `symmetry' conditions at first order, we may define functions $\tilde{B}_{2s}$ and $\tilde{B}_{2c}$ so that,
\begin{subequations}
    \begin{equation}
        B_{2s}=\tilde{B}_{2s}-\frac{1}{4}\left(1-\frac{1}{\eta_0'}\right)\frac{B_0^2d^2}{B_0'}\nu',
    \end{equation}
    \begin{multline}
        B_{2c}=\tilde{B}_{2c}+\frac{1}{8}\left(1-\frac{1}{\eta_0'}\right)\partial_\varphi\left(\frac{B_0^2d^2}{B_0'}\right)+\\
        +\frac{1}{8}\frac{\eta_0''}{(\eta_0')^2}\frac{B_0^2d^2}{B_0'}.
    \end{multline} 
\label{eqn:tildeDefs}
\end{subequations}
Written in this form, the QI conditions at second order simply become, 
\begin{subequations}
    \begin{gather}
        \tilde{B}_{2s}(\varphi)=\eta_0'(\varphi) \tilde{B}_{2s}[\eta_0(\varphi)], \\
        \tilde{B}_{2c}(\varphi)=\eta_0'(\varphi) \tilde{B}_{2c}[\eta_0(\varphi)].
    \end{gather} \label{eqn:B2tilde}
\end{subequations}
That is, omnigenity allows the second harmonics of the magnetic field to have a freely chosen part, so long as the choice satisfies the same symmetry as $d$ at first order. We have defined $\tilde{B}_{2s}$ and $\tilde{B}_{2c}$ in this particular form as it separates $B_{2s}$ and $B_{2c}$ naturally into two components with `opposite' symmetry; the explicit, lower order terms in Eqs.~(\ref{eqn:tildeDefs}) have the symmetry $f(\varphi)=-\eta_0'(\varphi)f[\eta_0(\varphi)]$. In other words, $B_{2s}$ and $B_{2c}$ each have a fixed even-parity part necessary to outweight the non-omnigeneous influence of $B_1$ at second order, as well as a free odd-parity part. 
\par
\subsection{QI conditions for the stellarator symmetric case} \label{sec:ssQInae}
We shall often be interested in learning about the reduced set of configurations that possess stellarator symmetry. Stellarator symmetry implies the invariance of $|\mathbf{B}|$ under a set of discrete, parity-like maps. Here we need only consider the symmetry around the minimum of the magnetic well, $(\theta,\varphi)\rightarrow(-\theta,-\varphi)$, which we take to be the symmetry point. This simplifies the QI conditions. 

\subsubsection{Symmetric magnetic well}
To leading order, stellarator symmetry implies,
\begin{equation}
    B_0(\varphi)=B_0(-\varphi).
\end{equation}
Then $\eta_0(\varphi)=-\varphi$, $\eta_0'=-1$ and $\eta_0''=0$. In the case of multiple wells stellarator symmetry will not generally simplify the problem for all minima.

\subsubsection{First order omnigenity}
At first order, stellarator symmetry requires,
\begin{equation}
    d(\varphi)\cos[\alpha+\nu(\varphi)]=d(-\varphi)\cos[-\alpha+\nu(-\varphi)].
\end{equation}
Using the omnigenity conditions, Eqs.~(\ref{eqn:omnCondeps}), and requiring the symmetry to hold at all field lines, then it follows that,
\begin{subequations}
    \begin{gather}
        d(\varphi)=-d(-\varphi), \\
        \nu(\varphi)=\frac{\pi}{2}+n\pi,
    \end{gather} \label{eqn:omnCondepsSS}
\end{subequations}
where $n\in \mathbb{Z}$. What is otherwise a function of the toroidal angle, $\nu$ is in the stellarator symmetric case required to be constant. Thus, $\nu'=0$ and the argument of the harmonic functions in Eq.~(\ref{eqn:Bform}) becomes simply the field line label $\alpha$.  The results to this order agree with those given in [\onlinecite{camacho-mata-2022}].

\subsubsection{Second order omnigenity}
Stellarator symmetry can straightforwardly be shown to require the functions $B_{20}$ and $B_{2c}$ to be even in $\varphi$, and $B_{2s}$ to be odd. Bringing these requirements together with the omnigenity conditions in Eqs.~(\ref{eqn:omnB2f}), 
\begin{subequations}
    \begin{gather}
        B_{20}(\varphi)=B_{20}(-\varphi), \label{eqn:omnCondeps2SS0}\\
        B_{2s}(\varphi)=-B_{2s}(-\varphi), \label{eqn:omnCondeps2SSs}\\
        B_{2c}(\varphi)=\frac{1}{4}\partial_\varphi\left(\frac{B_0^2d^2}{B_0'}\right). \label{eqn:omnCondeps2SSc}
    \end{gather}
\end{subequations}
Stellarator symmetry and the QI requirements coexist to leave $B_{20}$ and $B_{2s}$ to a large extent unconstrained, other than by their parity. The same is not true of $B_{2c}$. What in a general stellarator-symmetric stellarator would be a free even function, here is completely determined by the choice of the lower order functions $B_0(\varphi)$ and $d(\varphi)$. The reason is the clash between the symmetry requirements of stellarator symmetry and those of omnigenity. Thus, the choices at lower order will, through $B_{2c}$, affect properties such as shaping of flux surfaces and aspect ratio if the satisfaction of QI is sought also at second order.

\section{Pseudosymmetry and topological defects near field extrema} \label{sec:topDefects}
Getting to this point in second order we were not very cognizant of the behaviour of the expansion near points of $B_0'=0$. We saw that to leading order the behaviour at these points required $B_1$ to vanish wherever $B_0'$ did. However, we gained nothing concerning the behaviour of $B$ in the neighbourhood of these points. We know, though, that this is especially important for expressions like that of the perturbed map, $\eta_1$, Eq.~(\ref{eqn:eta1div}), as there could be a region of asymptotic breakdown around $B_0'=0$ where $\eta_1$ is not well behaved. Misbehaviour in this region could be seen to be a consequence of a change in $|\mathbf{B}|$ contours, and potentially, the change of their topology. Whether this occurs or not depends on the behaviour of $B_0$ and $B_1$ in the neighbourhood of the turning points. 
\par
\begin{figure}
        \centering
        \includegraphics[width=0.5\textwidth]{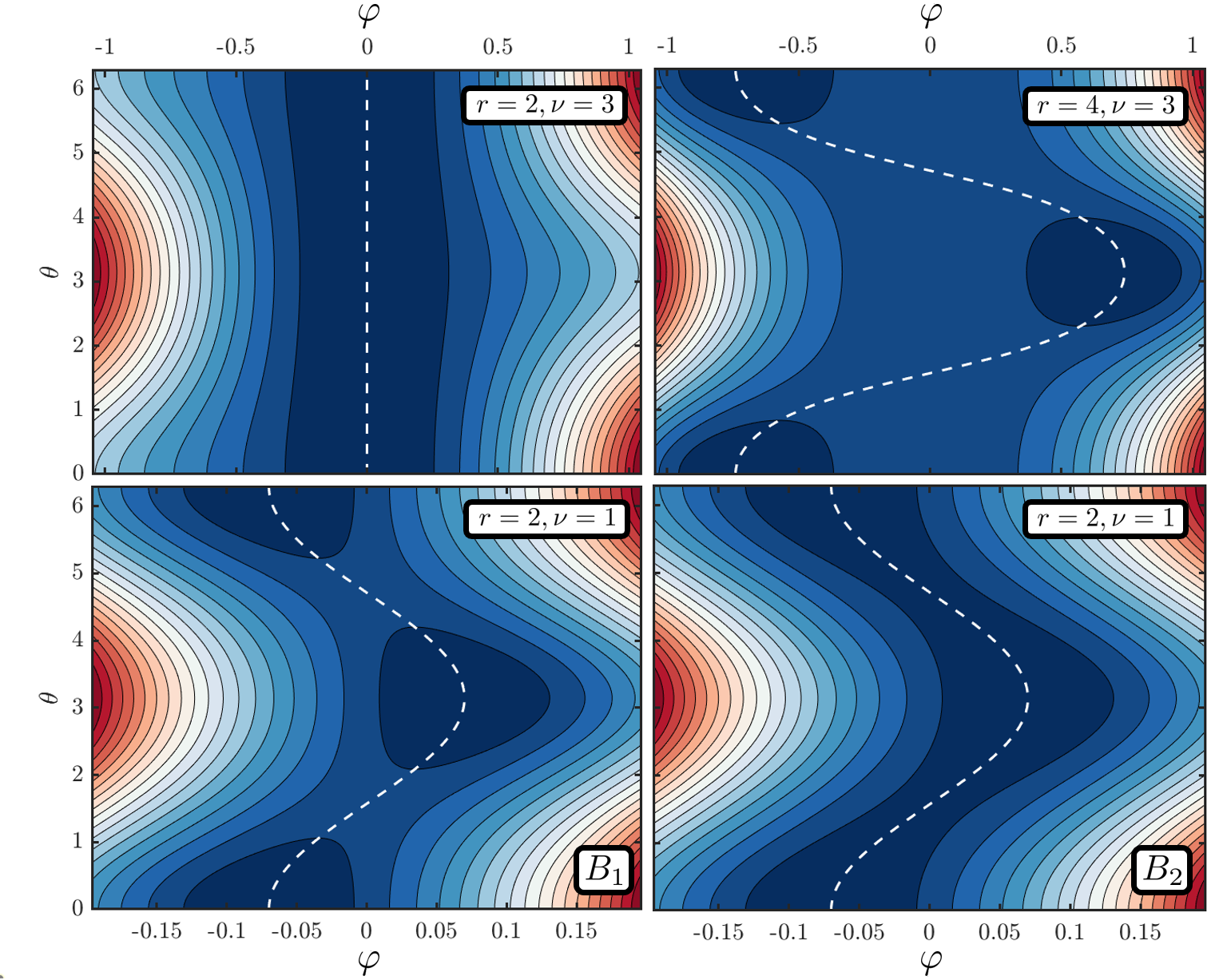}
        \caption{\textbf{Schematic picture of $|\mathbf{B}|$-contours with and without puddles.} Example of magnetic field contours with and without puddles. (i) First order $|\mathbf{B}|$ with $r=2,~\nu=3$, (ii) First order $|\mathbf{B}|$ with $r=2,~\nu=3$, (iii) First order $|\mathbf{B}|$ with $r=2,~\nu=1$ and (iv) Second order QI with $r=2,~\nu=1$. The broken white curve marks the minimum of $|\mathbf{B}|$ along lines of constant $\theta$.}
        \label{fig:Bcont}
\end{figure}

The magnetic field in the neighbourhood of these points may be modelled as $B_1\sim\varphi^\nu$ and $B_0'\sim\varphi^{r-1}$. We refer to the indices $\nu$ and $r$ as the order of the zeros of $B_1$ and $B_0$ respectively. When the perturbation $B_1$ is flat enough, $\nu\geq r$, then the extrema of $|\mathbf{B}|$ remain straight in the $(\theta,\varphi)$ plane, and the topology of the $|\mathbf{B}|$ contours is preserved (see Fig.~\ref{fig:Bcont}). In this convenient case the asymptotics in the neighbourhood of the turning point are correct. However, this choice of order of the zeroes is generally not a necessary condition for quasi-isodynamicity. In particular, this choice retains the straightness of $B_\mathrm{min}$, which we know is not implied by omnigenity. In fact, only the global maximum must be straight in QI configurations, but this conclusion only arises when periodicity is also considered\cite{cary1997,landreman2012}. To assess the behaviour near the turning points and assess the physical requirement on the order of the zeroes, we must bring the notion of pseudosymmetry onto the scene.
\par
\begin{figure}
    \centering \hspace*{-0.4cm}
    \includegraphics[width=0.5\textwidth]{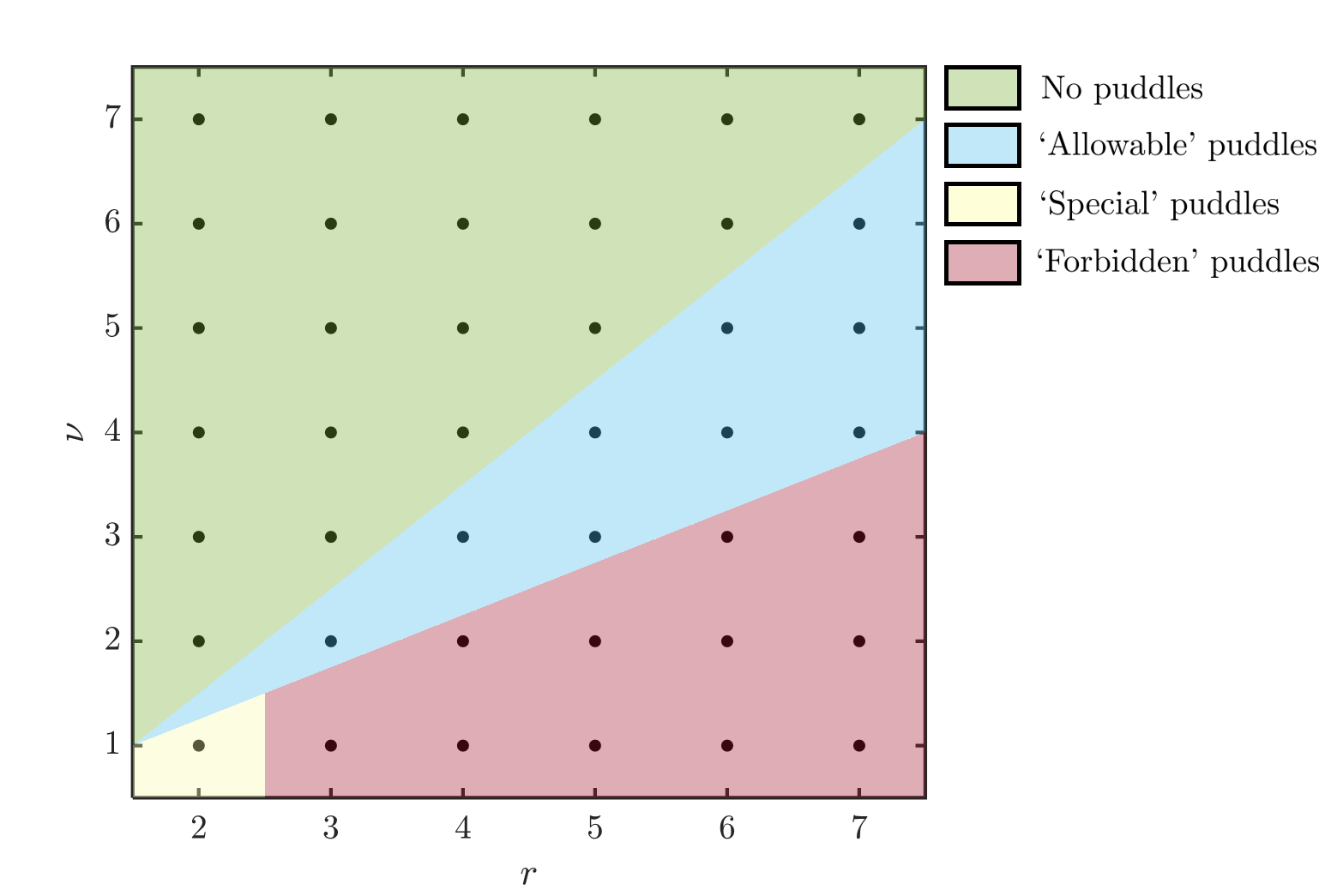}
    \caption{\textbf{Diagram of the order of zeroes and their implications.} The diagram shows the four distinct combinations of order of zeros of $\kappa$ and $B_0'$, $\nu$ and $r$ respectively, indicating no puddles in the contours of $|\mathbf{B}|$, `allowed', `special' ($r=2$ and $\nu=1$) and `forbidden' choices as determined by requiring radial drifts to vanish at turning points to second order in $\epsilon$.}
    \label{fig:Bdiag}
\end{figure}

For a magnetic field to be truly omnigeneous to the second order, the net drift of particles must be negligible to this order. Requiring the radial drift of particles to be of that order (or larger) at turning points of $|\mathbf{B}|$ (along field lines) provides several conditions on the field, the details of which may be found in Appendix~\ref{sec:appPseudo}. The behaviour near turning points depends on the values of the indices $\nu$ and $r$. These may be organised in four different categories (see Fig.~\ref{fig:Bdiag}) according to the presence of topological defects and whether pseudosymmetry (zero radial  drift at turning points) is achieved to order $\epsilon^2$. The condition that preserves the topology near extrema ($\nu\geq r$) is consistent with omnigeneity provided $\partial_\theta B_2=0$ at the turning points. To order $\epsilon^2$, defects on $|\mathbf{B}|$ contours arising from $2\nu>r>\nu>1$ preserve omnigeneity provided that $\partial_\theta B_2=0$ at the turning points of $B_0$. Those defects remain asymptotically small (higher order than $\epsilon^2$). 
\par
Not all topological defects described by the near-axis expansion are equally disruptive, though. All other combinations of $\nu$ and $r$, i.e. $\nu\leq r/2$, are too disruptive, with the exception of one very special case, perhaps the simplest. For $\nu=1$ and $r=2$, the choice of $B_{2c}$ consistent with QI, Eq.~(\ref{eqn:omnCondeps2SSc}), precisely cancels the leading disruptive contribution from the puddle. This marginal case is the only case in which second order can directly interact with the first order to amend omnigeneity. 
\par
Interestingly, all the scenarios in which the minimum of $|\mathbf{B}|$ is not straight (see white broken lines in Fig.~\ref{fig:Bcont}) do also exhibit topological defects in the form of puddles. The puddle size (which one may think of in terms of the variation of $B$ along the line of minima) scales like $\epsilon^2$ in the special case, and with a larger power in the case of the `allowable' puddles. In addition, only in the last special case does the $\varphi=0$ line not correspond to a contour of constant $|\mathbf{B}|$. The two lower examples of Fig.~\ref{fig:Bcont}, corresponding to this special case, demonstrate that the addition of the second order field $B_2$ can have the effect of ``healing'' the topological defects, in the sense that the large first-order (first harmonics in $\theta$) islands are broken and replaced by higher-order structures.  Although there is no hint that puddles can be completely eliminated at any finite order in $\epsilon$, this phenomenon resolves the apparent contradiction between the asymptotic construction of QI fields and the exact concept of QI fields with a non-straight minimum, as the latter absolutely forbids such topological features. This leaves open the possibility of an exactly QI solution near the bottom of the well.

\section{Conclusion} \label{sec:conclusions}
In this paper, we derive the conditions of quasi-isodynamicity on the magnetic field magnitude $|\mathbf{B}|$ near the magnetic axis. We do so by asymptotic expansion of the difference in radial drift at opposing portions of magnetic wells, providing a clear physical approach to the problem. This allows us to obtain QI conditions on the second order components of $|\mathbf{B}|$ in the distance from the magnetic axis.
\par
The approach and results in this paper set the ground for further exploration of quasi-isodynamic configurations and their properties using the near-axis framework, extending the original work in [\onlinecite{plunk2019}]. This includes consideration of appropriate shape choices, MHD stability, {\em etc}.
\par
Here, we have only considered the implications of omnigenity on the magnetic field magnitude. This is only part of the whole problem of constructing equilibrium fields, which includes also the full description of $\mathbf{B}$, requiring the solution of the so-called MHD constraint equations\cite{garrenboozer1991a,landreman2018a,rodriguez2020i}.  An analysis of the consequences of the second order QI conditions in this context will be presented in the future.

\section*{Acknowledgements}
The authors would like to acknowledge fruitful discussion with Per Helander and Rogerio Jorge.

\appendix
\section{On pseudosymmetry and topological defects in $B(\theta,\varphi)$} \label{sec:appPseudo}
In this appendix we consider the asymptotic considerations at and in the neighbourhood of turning points of $B_0$. Such turning points are special as deeply trapped and barely trapped particles spend an infinite amount of time at them. For the deeply trapped this is obvious, as these particles are unable to exist anywhere else. For the barely trapped it follows from $v_\parallel\sim\sqrt{1-(1-B_0''\varphi^2)}\sim\varphi$, and the divergence of $\int\mathrm{d}\varphi/v_\parallel\sim\log \varphi$ at the turning point. The consequence is that, to confine both classes of particles, we must make the radial drift \textit{exactly vanish} at those points. Formally, $\mathbf{B}\times\nabla B\cdot\nabla\psi=0$ wherever $\mathbf{B}\cdot\nabla B=0$. This is known as the condition of \textit{pseudosymmetry}\cite{mikhailov2002,skovoroda2005}.
\par
A magnetic field that is pseudosymmetric over a given magnetic flux surface, will possess contours of constant $|\mathbf{B}|$ all with the same topology\cite{landreman2012}. That is to say, the representation of $|\mathbf{B}|$ as a function of $(\theta,\varphi)$ should not present any contour that closes within a field period, i.e. features that resemble \textit{puddles}, which can be regarded as \textit{topological defects}. It is thus tempting to require such puddles not to be present at any of the asymptotic orders in which we have considered our QI construction. In the spirit of the asymptotic approach in this paper, though, it is only consistent to treat this pseudosymmetric condition asymptotically. This cannot be done through an asymptotic analysis of Eq.~(\ref{eqn:omnDefY}), both because the condition yields no information \textit{at} the turning point of $|\mathbf{B}|$, and the expansion itself breaks down in the neighbourhood. 
\par
Instead, to keep the behaviour at the extrema accountable, order by order, we shall assess the location of the extrema of $|\mathbf{B}|$ along field lines, and evaluate the radial drift there. Assessing the magnitude of the radial drift we may then deem the field consistent or inconsistent with omnigeneity (and thus also pseudosymmetry) to the right order. In practice, this requires an asymptotic expansion of both $\mathbf{B}\cdot\nabla B$ and $\mathbf{B}\times\nabla B\cdot \nabla \psi$. Fortunately, we already have these in Eqs.~(\ref{eqn:vddotdpsi}) and (\ref{eqn:BdotB}), as we needed them to construct $Y$. Thus all that remains, order-by-order, is (i) to find the turning points $\varphi_t=\varphi_t^{(0)}+\varphi_t^{(1)}+\dots$ and (ii) to evaluate the drift there.
\subsection{Order $\epsilon$}
To leading order $B=B_0(\varphi)$ and the extrema along field lines satisfy,
\begin{equation}
    B_0'(\varphi_t^{(0)})=0.
\end{equation}
The turning points of the magnetic field on axis define the turning points to leading order. The drift at these points is then, using Eq.~(\ref{eqn:vddotdpsi}),
\begin{equation}
    (I\partial_\varphi-G\partial_\theta)B\approx -\epsilon G_0 \partial_\theta B_1(\theta,\varphi_t^{(0)}).
\end{equation}
Note that this drift is order $\epsilon$, which is precisely the leading order of the drifts if the QI condition was not imposed. Thus, for a consistent choice to leading order, $B_1$ must vanish at all turning points. Note that the QI condition in Eq.~(\ref{eqn:B1omni}) only requires the vanishing of $B_1$ near minima (and the global maximum). The explicit confinement of barely trapped particles, though, requires it to vanish at all turning points. Therefore, locally $B_1(\theta,\varphi)\approx B_1^{(\nu)}(\theta)(\varphi-\varphi_t^{(0)})^\nu$, where $\nu\in\mathbb{N}$ is the order of the zero of $B_1$.

\subsection{Order $\epsilon^2$}
At the next order the location of the turning points must satisfy,
\begin{equation}
    B_0'(\varphi_t^{(0)}+\varphi_t^{(1)})+\epsilon(\partial_\varphi+\iota_0\partial_\theta)B_1(\theta,\varphi_t^{(0)}+\varphi_t^{(1)})=0.
\end{equation}
To solve this equation, we take $B_0'\approx B_0^{(r)}(\varphi-\varphi_t^{(0)})^{r-1}/(r-1)!$ about the turning point. The natural number $r>1$ must be even, and measures the shallowness of the magnetic well (or the top). We shall assume asymptotically $\varphi_t^{(1)}\sim O(\epsilon^\alpha)$ for $\alpha>0$ but not necessarily an integer. If this condition were not satisfied, then arbitrarily close to the axis the position of the extrema would be different from that defined by the axis. With this in mind, we may show that the only valid solutions to the equation are,
\begin{equation}
    \varphi_t^{(1)}=\begin{cases}
        0, & (\nu\geq r>1) \\
        0, \left[-\epsilon\frac{B_1^{(\nu)}}{B_0^{(r)}}\frac{(r-1)!}{(\nu-1)!}\right]^{1/(r-\nu)}, & (r>\nu>1) \\
        \left[-\epsilon\frac{B_1'}{B_0^{(r)}}(r-1)!\right]^{1/(r-1)}, & (\nu=1, r>1).
    \end{cases}
\end{equation}
Let us consider these possibilities in order.
\begin{itemize}
    \item \underline{$\nu\geq r>1$}: this scenario corresponds to one with a perturbation $B_1$ which is flatter than $B_0$. This yields a unique $|\mathbf{B}|$ extremum along $\mathbf{B}$, unchanged respect to the leading order. The radial drift at $\varphi=\varphi_t$ will to this order be,
    \begin{equation*}
        (I\partial_\varphi-G\partial_\theta)B\approx -\epsilon^2 G_0 \partial_\theta B_2(\theta,\varphi_t^{(0)}).
    \end{equation*}
    As in the previous order, for this drift to vanish to second order we must require $\partial_\theta B_2=0$. Note that this is not the same as $B_2=0$, as $B_2$ can have a non-zero $\theta$-average. Looking at the conditions on $B_2$ in the neighbourhood of the turning point, we see that (in stellarator symmetry) this is consistent with $B_{2s}$ being odd, and $B_{2c}\sim(\varphi^{2\nu}/\varphi^{r-1})'\sim\varphi^{2\nu-r}\sim0$. Such a field (see Fig.~\ref{fig:Bcont}) maintains the $\varphi=0$ line as a straight $|\mathbf{B}|$ contour. 
    
    \item \underline{$r>\nu>1$}: now consider the opposite case, in which $B_0$ is shallower than the correction $B_1$, leading to the possibility of multiple extrema. One remains at $\varphi=\varphi_t^{(0)}$ (where $B_1=0$), but another appears at a distance proportional to $\epsilon^{1/(r-\nu)}$. Its location will oscillate right and left of $\varphi=\varphi_t^{(0)}$, as the sign of $B_1^{(\nu)}(\theta)$ changes with the poloidal angle. This corresponds to the turning point, while $\varphi_t^{(0)}$ becomes an inflection point. The result is the change in the topology of $|\mathbf{B}|$ contours (see Fig.~\ref{fig:Bcont}), which may be quantified by the amount that $B$ changes along the line of minima. That is, $\delta B\sim\epsilon B_1^{(\nu)}(\varphi_t^{(1)})^\nu\sim\epsilon^{1+\nu/(r-\nu)}$, which is of an order higher than 2 for $2\nu>r$.
    \par
    The drift behaviour at $\varphi=\varphi_t^{(0)}$ requires $\partial_\theta B_2=0$ as in the previous case, consistent with the QI conditions in its neighbourhood. To assess the implications of the additional turning point, let us evaluate the drift,
    \begin{multline*}
        (I\partial_\varphi-G\partial_\theta)B\approx-\epsilon G_0\partial_\theta B_1^{(\nu)}\frac{(\varphi_t^{(1)})^\nu}{\nu!}+\\
        +\epsilon^2\left[I_2B_0^{(r)}\frac{(\varphi_t^{(1)})^{r-1}}{(r-1)!}-G_0\partial_\theta B_2^{(t)}\frac{(\varphi_t^{(1)})^t}{t!}\right],
    \end{multline*}
    where we took $\partial_\theta B_2\propto(\varphi-\varphi_t^{(0)})^t$, and $t\in\mathbb{N}$. The correction $\varphi_t^{(1)}\propto \epsilon^{1/(r-\nu)}$, so the three terms in the drift involve the following powers of $\epsilon$ respectively: $r/(r-\nu)$, $(r-1)/(r-\nu)+2$ and $2+t/(r-\nu)$. The second term is always subdominant to the first, as $r-\nu\geq 1$. The first term dominates if the $B_2$ zero is of high enough order, $t>2\nu-r$. Because there is no way of making such a term vanish (as by assumption $\partial_\theta B_1^{(\nu)}$ is non-vanishing, at least for some $\theta$), then the only option left to enforce pseudosymmetry to the appropriate order is to make the order of this term large enough, namely $r/(r-\nu)>2$. This requires $2\nu>r$, that is, the order of the zero of $B_1$, which is by assumption smaller than that of $B_0$, not to be too small. To make it order $\epsilon^k$, $k\nu/(k-1)>r>\nu$. In the case of $t\leq2\nu-r$ the $B_2$ term is dominant (in the equal case of the same order as the first term), but always gives a power of $\epsilon$ that is greater than 2. Thus, the deviation from pseudosymmetry is higher order. In this case as well, $2\nu>r$ for $t\neq0$.
    \par
    In summary, we have a second possibility which allows for topological defects in $|\mathbf{B}|$ but avoids large particle losses so long as,
    \begin{equation}
        2\nu>r>\nu>1.
    \end{equation}
    Note that the appearance of these puddles makes the asymptotic approach for the QI behaviour in the main text fail in the neighbourhood of the extrema for $r>\nu+1$. This is indicated by the divergence of $\eta_1$, which is expected given the movement of the minimum and thus the non-smooth change in the bounce map definition.
    
    \item \underline{$\nu=1, r\geq2$}: in this special case there is a single turning point displaced from $\varphi=\varphi_t^{(0)}$. The turning point obeys $\varphi_t^{(1)}\propto \epsilon ^{1/(r-1)}$, which makes the three terms in the drift have the following powers of $\epsilon$: $r/(r-1)$, $3$ and $2+t/(r-1)$, where $t$ can in principle be zero here, as there is no additional requirement stemming from $\varphi_t^{(0)}$.
    \par
    In the case of $r>2$, the power of the first term becomes smaller than $3/2$, and thus dominate over the second and last terms. This makes, asymptotically, omnigeneity to be broken at second order, and thus this form cannot be allowed. The special case that remains to consider is $r=2$ and $\nu=1$. In that case, for $t=0$, the first and last terms are both order $\epsilon^2$. These terms may therefore compete with each other at $O(\epsilon^2)$, to vanish if
    \begin{equation}
        \partial_\theta B_2(\theta,\varphi_t^{(0)})=\partial_\theta\left(\frac{(B_1')^2}{2B_0''}\right)(\theta,\varphi_t^{(0)}).
    \end{equation}
    This balance is precisely of the form enforced by the QI requirement in Eq.~(\ref{eqn:omnCondeps}), which suggests that it is possible, in principle, to take $\nu=1$ and $r=2$.
\end{itemize}
In summary, then, to second order, the pseudosymmetry condition requires one of the following three,
\begin{enumerate}
    \item for $\nu\geq r$: $\partial_\theta B_2(\theta,\varphi_t^{(0)})=0$, which preserves the topology of the contours of $|\mathbf{B}|$ to this order.
    \item for $2\nu>r>\nu>1$: $\partial_\theta B_2(\theta,\varphi_t^{(0)})=0$, breaks the topology of $|\mathbf{B}|$-contours with the appearence of puddles, but the derived break-down of omnigeneity is higher order. 
    \item for $r=2,~\nu=1$: $\partial_\theta B_2=\partial_\theta((B_1')^2/2B_0'')$ at $(\theta,\varphi_t^{(0)})$. This is satisfied by the QI conditions to second order around the bottoms of the wells. It also gives puddles. 
\end{enumerate}

The above consideration gives a sense of the importance of the turning points, and the behaviour of the various near-axis functions about them. We saw that in certain cases, the QI conditions derived in this paper do not apply close to the turning points. An example of that was the $2\nu>r>\nu>1$ case. Misbehaviour near the minimum affects not only the deeply trapped population, but also the remainder classes, which must physically traverse the region. To estimate by how much, consider that some region $\Delta\varphi\sim\epsilon^\alpha$ is spoiled near the minimum. And in that region, the drift to be order $\epsilon^\beta$. Then, we expect the effect on the bounce averaged drift to be $O(\epsilon^{\alpha+\beta})$. In the $2\nu>r>\nu>1$ situation (with $t$ large enough), $\alpha\sim 1/(r-\nu)$ and $\beta\sim r/(r-\nu)$. Thus the spoiling of QI will be order $(r+1)/(r-\nu)\geq2+2/(\nu-1)>2$, limit in which $r=2\nu-1$ (the lower limit of $r=\nu+1$ gives an order of three or larger). 
\par

\section{Derivation of second order theory using approach of \citet{plunk2019}}
\label{sec:appCS}
Here we sketch an alternative derivation of the second order omnigenity condition using the approach based on bounce distance invariance, {\em i.e.} the property of equal distance between bounce points \cite{plunk2019, camacho-mata-2022}.  This is the principle underlying the constructive form of the omnigenity condition\cite{cary1997} which makes use of a coordinate $\eta$ (not to be confused with the bounce function used here in the main text), in which the contours of the magnetic field strength appear straight.\footnote{A minimal formulation can be made of this equal-distance criterion as follows. Define the $|\mathbf{B}|$-contour label $\eta$ through $\bar{B}(\eta)=B(\theta,\varphi)$, which also introduces the function $\bar{B}$. Then, the equal distance criterion is $B(\theta,\varphi)=B(\theta+\iota\Delta\varphi,\varphi+\Delta\varphi)$, where $\Delta\varphi$ is a function only of $\eta$, and it meadures the angular separation between bounce points. Expanding $\eta$, $\bar{B}$, $B$ and $\Delta\varphi$, the same conditions as in the appendix are reached. }  For QI fields, $\eta$ is defined by the mapping

\begin{equation}
    \varphi = \eta + F(\theta, \eta),
\end{equation}
where the omnigenity condition can be expressed as a symmetry on $F$,\cite{plunk-nss-2023}
\begin{equation}
    F(\theta, \eta) = F(\theta - \iota \Delta\eta, \eta_b)
\end{equation}
and the magnetic field strength can be expressed as a function only of $\eta$, $B = \bar{B}(\eta)$.  The quantities $\Delta\eta$ and $\eta_b$ are functions defined by
\begin{equation}
\bar{B}(\eta_b(\eta)) = \bar{B}(\eta),
\end{equation}
with the trivial root $\eta_b(\eta) = \eta$ excluded except at the minimum of $\bar{B}$.  The angular distance is then defined $\Delta\eta(\eta) = \eta - \eta_b(\eta)$.  We expand (``near quasi-symmetry'')
\begin{eqnarray}
\eta = \eta_0 + \epsilon \eta_1 + \epsilon^2  \eta_2 + \dots,\\
F = \epsilon F_1 + \epsilon^2 F_2 + \dots,\\
\bar{B} = \bar{B}_0 + \epsilon^2 \bar{B}_2 + \dots,
\end{eqnarray}
{\em i.e.} $F_0 = 0$ and at dominant order $\eta_0 = \varphi$, and the zeroth order magnetic field is $B_0(\varphi) = \bar{B}_0(\varphi)$.  Due to the freedom in defining $\eta$, we need not perturb functions $\eta_b$, and $\Delta\eta$.  At first order we find
\begin{eqnarray}
    \eta_1 = -F_1(\theta, \varphi),\\
    F_1(\theta, \varphi) = F_1(\theta - \iota\Delta\varphi, \varphi_b)
\end{eqnarray}
where we have defined the more physically transparent notation $\varphi_b(\varphi) = \eta_b(\eta_0) = \eta_b(\varphi)$, $\Delta \varphi = \varphi - \varphi_b$.  From the expansion of $\bar{B}$, we have $B_1 = \eta_1 B_0^\prime(\varphi)$, and therefore we can obtain the equivalent of Eqn.~\ref{eqn:B1omni}, namely

\begin{equation}
    B_1(\theta, \varphi) = \varphi_b^\prime B_1(\theta - \iota \Delta\varphi, \varphi_b)
\end{equation}
where functions of $\varphi_b^\prime(\varphi) = \varphi_b^\prime$, {\em etc.}, for succinctness. It is worth noting that the relationship between $B_1$ and $\eta_1$ requires that $B_1$ is zero at all extrema (both minima and maxima) in order for the coordinate mapping $\eta_1$ to be well behaved. Thus pseudosymmetry is encoded in this approach, at least at first order. At next order we find 

\begin{eqnarray}
    \eta_2 = -F_2(\theta, \varphi) - \eta_1 \partial_\varphi F_1(\theta, \varphi),\label{eq:eta2}\\
    F_2(\theta, \varphi) = F_2(\theta - \iota \Delta \varphi, \varphi_b).\label{eq:F2}
\end{eqnarray}
The second order contributions to $B$ are $B_2(\theta, \varphi) = \eta_1^2 B_0^{\prime\prime}/2 + \eta_2 B_0^{\prime} + \bar{B}_2(\varphi)$.  Combining this with Eqns.~\ref{eq:eta2}-\ref{eq:F2}, we obtain a symmetry condition on $B_2$ equivalent to Eqn.~\ref{eqn:B2omni},
\begin{widetext}
    \begin{equation}
        B_2(\theta, \varphi) - \bar{B}_2(\varphi) - \partial_\varphi\left(\frac{B_1^2}{2B_0^\prime} \right) = \varphi_b^\prime \left[ B_2(\theta, \varphi) - \bar{B}_2(\varphi) - \partial_\varphi\left(\frac{B_1^2}{2B_0^\prime} \right) \right]_b
    \end{equation}
\end{widetext}
where we have introduced a notation to signify the application of replacement rules: $[A]_b = A_{\theta \rightarrow \theta - \iota \Delta\varphi, \varphi \rightarrow \varphi_b}$.

Finally, for more direct comparison with Eqn.~\ref{eqn:omnB2f}, we can rewrite this in a form where quantities are explicitly evaluated at $\theta$ and $\varphi$

\begin{widetext}
    \begin{equation}
        B_2(\theta, \varphi) - \bar{B}_2(\varphi) = \varphi_b^\prime\left[ B_2(\theta - \iota \Delta\varphi, \varphi_b) - \bar{B}_2(\varphi_b) \right] + \left(\frac{B_1^2}{2B_0'}\right)\frac{\varphi_b^{\prime\prime}}{(\varphi_b^\prime)^2} + \frac{1}{2}\left(1-\frac{1}{\varphi_b^\prime}\right)(\partial_\varphi+\iota \partial_\theta )\left(\frac{B_1^2}{B_0^\prime}\right).
    \end{equation}
\end{widetext}
We note that the symmetry conditions here do not appear under the derivative $\partial_\theta$, as they are in some sense integrated versions of Eqns.~(\ref{eqn:B1omni}) and (\ref{eqn:B2omni}).  Also for this reason the second order condition contains the $\theta$-independent free function $\bar{B}_2$.  As we have argued, however, there is no constraint related to omnigenity that needs to be satisfied on the $\theta$ dependent part of $B_2$.

\bibliography{condQS}

\section*{Data availability}
Data sharing is not applicable to this article as no new data were created or analyzed in this
study.

\end{document}